\documentclass[lettersize,journal]{IEEEtran}
\IEEEoverridecommandlockouts
\usepackage{hyperref}
\usepackage{url}
\usepackage{tikz}
\usepackage{caption}
\usetikzlibrary{positioning, shapes.geometric, arrows.meta, fit}
\usetikzlibrary{shapes.geometric, arrows.meta}
\tikzstyle{blockish} = [rectangle, minimum width=3cm, minimum height=1cm, text centered, draw=black, fill=blue!30]
\tikzstyle{startstop} = [rectangle, rounded corners, minimum width=3cm, minimum height=1cm,text centered, draw=black, fill=red!30]
\tikzstyle{process} = [rectangle, minimum width=3cm, minimum height=1cm, text centered, draw=black, fill=blue!30]
\tikzstyle{decision} = [diamond, minimum width=3cm, minimum height=1cm, text centered, draw=black, fill=green!30]

\tikzstyle{block} = [rectangle, rounded corners, minimum width=2.5cm, minimum height=1cm, text centered, draw=black, fill=green!30]
\tikzstyle{block2} = [rectangle, rounded corners, minimum width=2.5cm, minimum height=1cm, text centered, draw=black, fill=blue!30]
\tikzstyle{block3} = [rectangle, rounded corners, minimum width=2.5cm, minimum height=1cm, text centered, draw=black, fill=red!30]
\tikzstyle{block_start_end} = [rectangle, rounded corners, minimum width=2.5cm, minimum height=1cm, text centered, draw=black, fill=gray!30]
\tikzstyle{arrow} = [thick,->,>=stealth]

\usepackage{cite}
\usepackage{balance}
\usepackage{comment}
\usepackage{amsmath,amssymb,amsfonts}
\usepackage{algorithmic}
\usepackage{tcolorbox}
\usepackage{graphicx}
\usepackage{multirow}
\usepackage{amsmath}
\usepackage{booktabs}
\usepackage{textcomp}
\usepackage{xcolor}
\def\BibTeX{{\rm B\kern-.05em{\sc i\kern-.025em b}\kern-.08em
    T\kern-.1667em\lower.7ex\hbox{E}\kern-.125emX}}
    
\begin{document}

\title{What People Share With a Robot When Feeling Lonely and Stressed and How It Helps Over Time}

\author{Guy Laban$^{1,*}$, Sophie Chiang$^{1}$, Hatice Gunes$^{1}$\\
{
    $^1$ Department of Computer Science and Technology, University of Cambridge, Cambridge, United Kingdom\\
}
\thanks{{Corresponding author: Guy Laban \tt\small {guy.laban}@cl.cam.ac.uk}}%
\thanks{G. Laban and H. Gunes are supported by the EPSRC project ARoEQ under grant ref. EP/R030782/1.} \thanks{\textbf{Open Access:} For open access purposes, the authors have applied a Creative Commons Attribution (CC BY) licence to any Author Accepted Manuscript version arising. \textbf{Data access:} Raw data related to this publication cannot be openly released due to anonymity and privacy issues.}
}

\maketitle

\begin{abstract}
Loneliness and stress are prevalent among young adults and are linked to significant psychological and health-related consequences. Social robots may offer a promising avenue for emotional support, especially when considering the ongoing advancements in conversational AI. This study investigates how repeated interactions with a social robot influence feelings of loneliness and perceived stress, and how such feelings are reflected in the themes of user disclosures towards the robot. Participants engaged in a five-session robot-led intervention, where a large language model powered QTrobot facilitated structured conversations designed to support cognitive reappraisal. Results from linear mixed-effects models show significant reductions in both loneliness and perceived stress over time. Additionally, semantic clustering of 560 user disclosures towards the robot revealed six distinct conversational themes. Results from a Kruskal-Wallis H-test demonstrate that participants reporting higher loneliness and stress more frequently engaged in socially focused disclosures, such as friendship and connection, whereas lower distress was associated with introspective and goal-oriented themes (e.g., academic ambitions). By exploring both how the intervention affects well-being, as well as how well-being shapes the content of robot-directed conversations, we aim to capture the dynamic nature of emotional support in human–robot interaction.
\end{abstract}

\section{Introduction}

Loneliness and stress are growing concerns among young adults \cite{Arnett2014TheHealth, Shah2023UnderstandingInterventions.}. Persistent feelings of disconnection and psychological strain impair well-being while also being linked to a range of negative health outcomes, including depression and anxiety 
\cite{Arnett2014TheHealth}. While several interventions aim to reduce loneliness and stress \cite{Magid2023TheStudy, Ellard2023Review:Review, Masi2010ALoneliness}, much less is known about how these emotional states evolve during repeated interactions with a social robot, and how they might shape, or be shaped by the content of such interactions. Social robots, embodied agents that are aimed at interacting socially with humans \cite{breazeal_designing_2004}, offer a promising avenue for supporting emotional well-being \cite{Laban2024SocialWell-Being}. Their physical presence, responsiveness, and capacity to engage in relational dialogue position them as compelling tools for social interaction \cite{Henschel2021}, as well as delivering socio-emotional support \cite{RefWorks:404}. Prior work has shown that robots can foster self-disclosure \cite{Laban2024SharingFeel,Laban2021TellSpeech}, elicit trust \cite{Langer2019,Stower}, and play a role in both therapeutic and educational contexts \cite{Henschel2021}. 

While previous studies have demonstrated the potential of robot-led interventions to reduce feelings of loneliness and stress \cite{Laban2024BuildingTime,laban_ced_2023}, there remains a pressing need for further empirical evidence. This need is particularly important in light of ongoing advancements in conversational AI , with social robots now leveraging large language models (LLMs) to offer richer, more socially engaging experiences that may have a more significant impact on users' emotional wellbeing \cite{SpitaleMicol2025VITA:Coaching,Laban2025AReappraisal}. Accordingly, we are asking: \textit{\textbf{(RQ1)} To what extent does interacting with a robot influence people's feelings of loneliness and perceptions of stress?}

Furthermore, 
little is known about the content of these conversations, specifically, what individuals choose to discuss in such interactions when experiencing heightened loneliness or stress. Gaining insight into these conversational choices is crucial for understanding how people seek emotional support from robots and for designing interventions that are more attuned to users’ actual needs and concerns. Accordingly, we are asking: \textit{\textbf{(RQ2)} To what extent are different themes of self-disclosure during repeated interactions with a robot are associated with variations in loneliness and stress?}

This study takes a dual approach to these questions. First, we assess whether engaging in a longitudinal robot-led intervention, aimed at supporting cognitive reappraisal \cite{Laban2025AReappraisal}, can reduce participants’ reported loneliness and perceived stress. Second, we examine how such feelings are reflected in the topics they choose to disclose to the robot. By exploring both how the intervention affects well-being, as well as how well-being shapes the content of robot-directed conversations, we aim to capture the dynamic nature of emotional support in human–robot interaction (HRI).


\section{Related works}
\label{related}


Loneliness and stress are often co-occurring emotional states that significantly impact psychological and physical well-being \cite{Cacioppo2006LonelinessPerspective}. While loneliness is typically defined as the subjective experience of social isolation or the discrepancy between desired and actual social connections \cite{Hawkley2010LonelinessMechanisms}, stress is a broader physiological and emotional response to perceived threats or challenges \cite{lazarus1984stress}. Both states trigger distinct yet overlapping behavioural and conversational patterns, especially during moments of self-expression and interpersonal sharing \cite{Delgado2023CharacterizingConnection}. When individuals experience loneliness, their conversations often reflect unmet social needs. People tend to talk about feelings of being disconnected, left out, or misunderstood. They may revisit past relationships, express longing for companionship, or speak about the absence of meaningful social roles \cite{Hawkley2010LonelinessMechanisms,Horowitz1979InterpersonalLonely}. 
Stress-related conversations are typically more focused on situational pressures. People experiencing stress frequently talk about sources of overwhelm, such as academic or professional demands, financial worries, interpersonal conflict, or health concerns. Stress talk is often future-oriented, with individuals anticipating consequences or strategising solutions \cite{Pennebaker1993PuttingImplications}. 
Despite these distinctions, loneliness and stress often intersect in conversations, particularly in narratives about lacking support. Individuals may articulate stress over life demands while simultaneously expressing the absence of someone to lean on, suggesting that social isolation amplifies perceived stressors \cite{Hawkley2010LonelinessMechanisms,Delgado2023CharacterizingConnection,Cacioppo2014SocialIsolation, Cacioppo2011SocialIsolation}. 
Previous studies explain that interpersonal communication, especially when characterized by empathy, active listening, and emotional validation, can significantly alleviate feelings of loneliness and stress by fostering a sense of connection, shared understanding, and psychological safety \cite{Segrin2007PositiveWell-being, Segrin2010FunctionsHealth, Zaki2013InterpersonalRegulation,Zaki2020IntegratingRegulation}.


Previous studies also explored the roles of social robots in supporting individuals with loneliness and stress, and how such feelings might effect individuals' interactions with social robots \cite{Gasteiger2021FriendsPeople, Rasouli2022PotentialAnxiety, Laban2022SocialTreatmentd}. Previous research has found that individuals tend to disclose more to social robots when experiencing higher levels of loneliness and stress \cite{2023OpeningBehavior}. This has been further supported by cross-sectional research, showing that individuals who experienced situational loneliness during the COVID-19 pandemic reported being particularly prone to accepting social robots (when presented with an image of the NAO robot) as social interaction partners \cite{penner2022}. These findings align with prior studies indicating that people experiencing loneliness \cite{RefWorks:300,Epley2008CreatingArticle,Yamaguchi2023YoungStress} and stress \cite{Yamaguchi2023YoungStress} are more likely to anthropomorphise. A recent study found that individuals who felt lonelier ascribed significantly more social characteristics to a non-humanoid robot and were more likely to exhibit social behaviour towards it \cite{Leichtmann2025LonelyRobots}. In the context of conversational interactions with social robots, one study reported that after a 10-session (five-week) intervention involving pre-scripted conversations with the social robot Pepper, participants communicated more over time and reported feeling less lonely and in a better mood both over time and after each session \cite{Laban2024BuildingTime}. A replication of the study with a population prone to emotional distress (i.e., informal caregivers) yielded similar results, additionally showing a positive effect on perceived stress. Participants also reported being more accepting of their caregiving situation, reappraising it more positively, and experiencing less blame towards others \cite{laban_ced_2023}.

As social robots become increasingly adaptive through the use of LLMs \cite{10715872}, their ability to simulate responsive communication can foster a sense of rapport. However, this same level of adaptivity may blur the boundaries of perceived confidentiality, potentially affecting the emotional outcomes of such interactions \cite{Laban2024SharingFeel}. Therefore, it is important to provide further empirical evidence on how repeated conversational interactions with a social robot, specifically one powered by an LLM and designed to adapt to users’ input and affect people’s emotions and feelings. 
Moreover, existing research has not examined how the evolving content of users’ disclosures during repeated interactions reflects or relates to their loneliness and perceived stress. This study addresses that gap by exploring the relationship between these feelings and conversational themes in longitudinal robot-led interactions.

\section{Methods}

\subsection{Dataset}

\begin{figure}[h]
    \centering
    \includegraphics[width=.49\columnwidth]{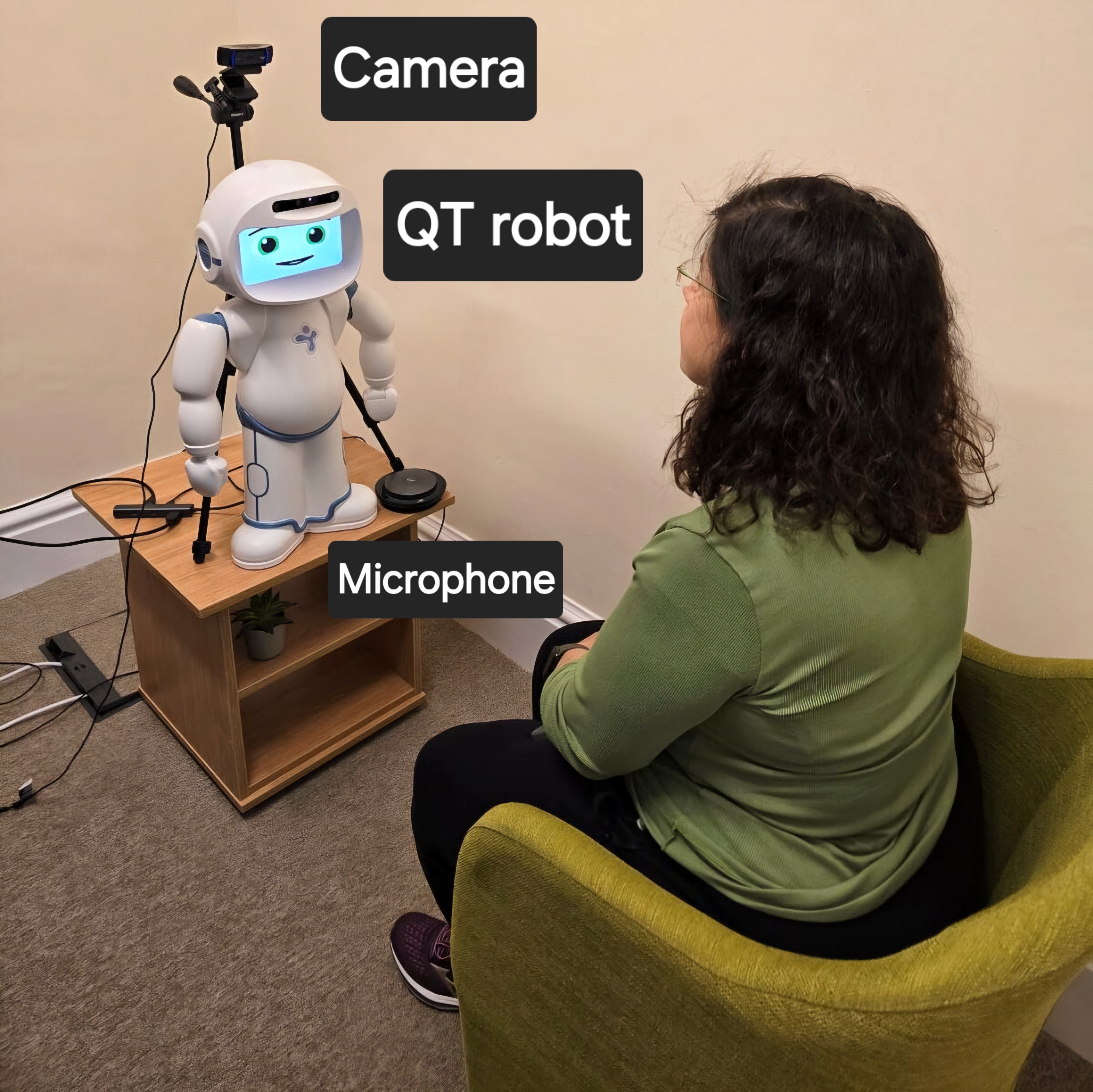} 
    \includegraphics[width=.49\columnwidth]{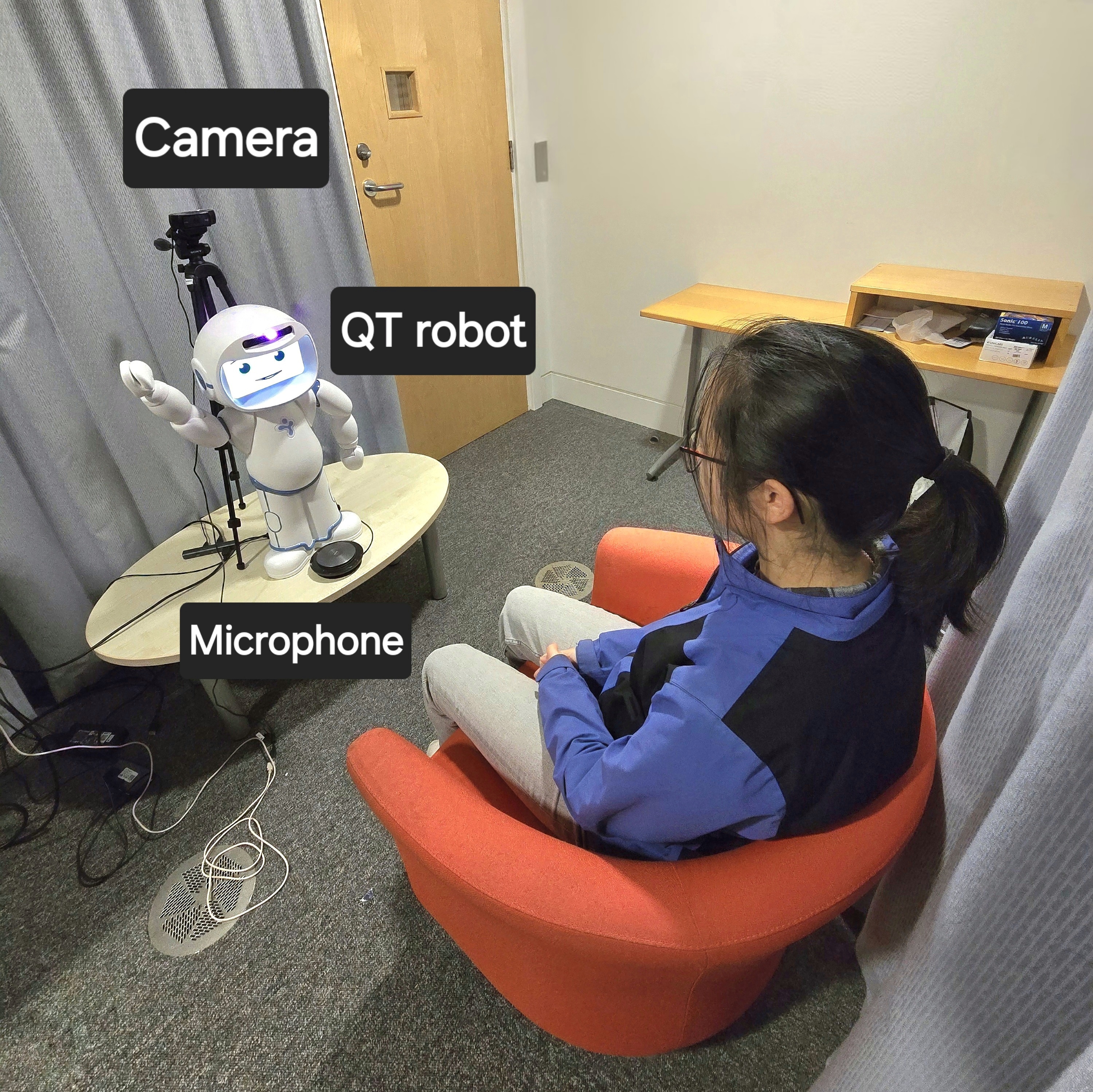}
    \caption{\footnotesize The deployment settings. Image from \cite{Laban2025AReappraisal}.}
    \label{fig:set}
\end{figure}

To address our research questions, we used data collected and reported in \cite{Laban2025AReappraisal}. Data were collected through a five-session robot-led intervention conducted with 21 university students in familiar settings such as university halls and departments (see Figure \ref{fig:set}). Each participant engaged in a structured intervention that contained conversations with the robot ``QTrobot" (LuxAI), which facilitated cognitive reappraisal using a LLM (GPT-3.5), by utilising an open-sourced robotic system \cite{SpitaleMicol2025VITA:Coaching}. The sessions followed the PERMA framework \cite{Seligman2018PERMAWell-being}, with two positive-connotation and one negative-connotation question per session, where the robot guided participants to reinterpret emotionally charged experiences. Data collection included standardized self-report questionnaires as well as interaction logs of the disclosures' content towards the robot. The study was approved by the departmental ethics committee and participants provided informed consent before participating in the study. For more information about the data collection methodology, see \cite{Laban2025AReappraisal}.

\subsection{Measurements}
\label{measures}

\paragraph{Loneliness} After the first and the last sessions, participants were asked to report their feelings and thoughts of loneliness over the previous three days using the short-form UCLA loneliness scale (ULS-8; see\cite{Hays1987}). The scale includes 8 items rated on a seven-point scale, ranging from 1 (not at all) to 7 (all the time). Accordingly, a mean scale was constructed (\(M\) = 4.80, \(SD\) = 1.13) which was found to be reliable (Cronbach's \(\alpha\) = .89). 
\paragraph{Stress} After the first and the last sessions, participants were asked to report their feelings and thoughts of periodic stress from the past month using the perceived stress scale \cite{RefWorks:475}. The scale includes 10 statement items rated on a seven-point scale, ranging from 1 (never) to five (very often). A mean scale was constructed (\(M\) = 3.96, \(SD\) = 0.89) which was found to be reliable (Cronbach's \(\alpha\) = .87).


\subsection{Data Analysis}
\label{data_ana}

\subsubsection{The intervention effect on Loneliness and Stress}

Linear mixed-effects models (LME) were employed using  lme4 for R \cite{Bates2015FittingLme4} to evaluate repeated measures between the first and the fifth sessions, accounting for both fixed effects (the session number) and random effects (accounting for participant variance). These models were selected for their ability to manage 
repeated-measures data and ensure robustness in capturing individual differences \cite{Bell2019FixedChoice}. Additionally, they were well-suited for data from two cohorts, as modelling participants as random effects inherently accounts for individual variability within and across cohorts, ensuring that cohort-level differences do not confound the overall analysis \cite{Hesser2015ModelingInterventions}. Significance was calculated using lmerTest \cite{Kuznetsova2017LmerTestModels} applying Satterthwaite’s method \cite{Satterthwaite1946AnComponents}. 

\subsubsection{Assessing the variations in loneliness and stress based on disclosure theme}

Each disclosure towards the robot was first preprocessed by removing stopwords after splitting the text by whitespace. 
The resulting texts were then encoded into 384-dimensional sentence embeddings using the all-MiniLM-L6-v2 model \cite{10.5555/3495724.3496209} from the SentenceTransformers library \cite{Reimers2019Sentence-BERT:BERT-Networks}. These embeddings were clustered using the K-means algorithm \cite{likas2003global}, with a fixed random seed to ensure replicability. The optimal number of clusters was determined using the elbow method based on inertia values \cite{liu2020determine}. To generate human-readable labels and explanations for each cluster, the top $n$ responses nearest to each cluster centroid were concatenated and passed as a prompt to GPT-4o-mini. The model was instructed to output a concise cluster label and a detailed description of key themes using the 
prompt:

\begin{tcolorbox}
\textit{\small “The following are responses to questions from a specific cluster. Analyse these responses and provide: 1) A concise label summarizing the main theme or central topic of this cluster; 2) A detailed paragraph describing key themes, patterns, or insights. Highlight any notable 
trends specific to this cluster.
Do not include any introductory statements or additional commentary.
Only provide the label and description, without introductory statements or commentary.”}
\end{tcolorbox}

As a sanity check, the semantic validity of each LLM-generated description was evaluated using cosine similarity between the embedding of the description and its corresponding cluster centroid, computed using the same all-MiniLM-L6-v2 model \cite{10.5555/3495724.3496209}.

To assess how different topics of interaction with the robot were associated with varying levels of reported loneliness and perceived stress, non-parametric statistical tests were employed due to the non-normal distribution of outcome variables. A Kruskal-Wallis H-test was conducted separately for loneliness and perceived stress to assess differences across the generated clusters. Upon identifying significant omnibus effects, pairwise post-hoc comparisons were conducted using Mann-Whitney U tests to explore specific differences between cluster pairs. All pairwise comparisons were two-tailed and unadjusted to preserve sensitivity in identifying meaningful differences in this exploratory analysis. 
This approach enabled us to examine how the topic of participants disclosures to the robot may reflect their feelings of loneliness and stress.

\section{Results}

\subsection{The effect of the intervention on Loneliness}

The intervention with QT had a significant positive effect on participants’ feelings of loneliness. 
As demonstrated in Figure \ref{fig:pss}, the results emphasize that despite the variance between participants (±0.98), participants’ loneliness scores significantly decreased over time, indicating reduced feelings of loneliness, $\beta$ = –0.15, $SE$ = 0.01, $p$ $<$ .001 (see Table \ref{tab:lon_pss}). This suggests that repeated interactions with QT were associated with decreased loneliness across sessions.

\subsection{The effect of the intervention on perceived stress}

\begin{figure}
    \centering
    \includegraphics[width=0.49\linewidth]{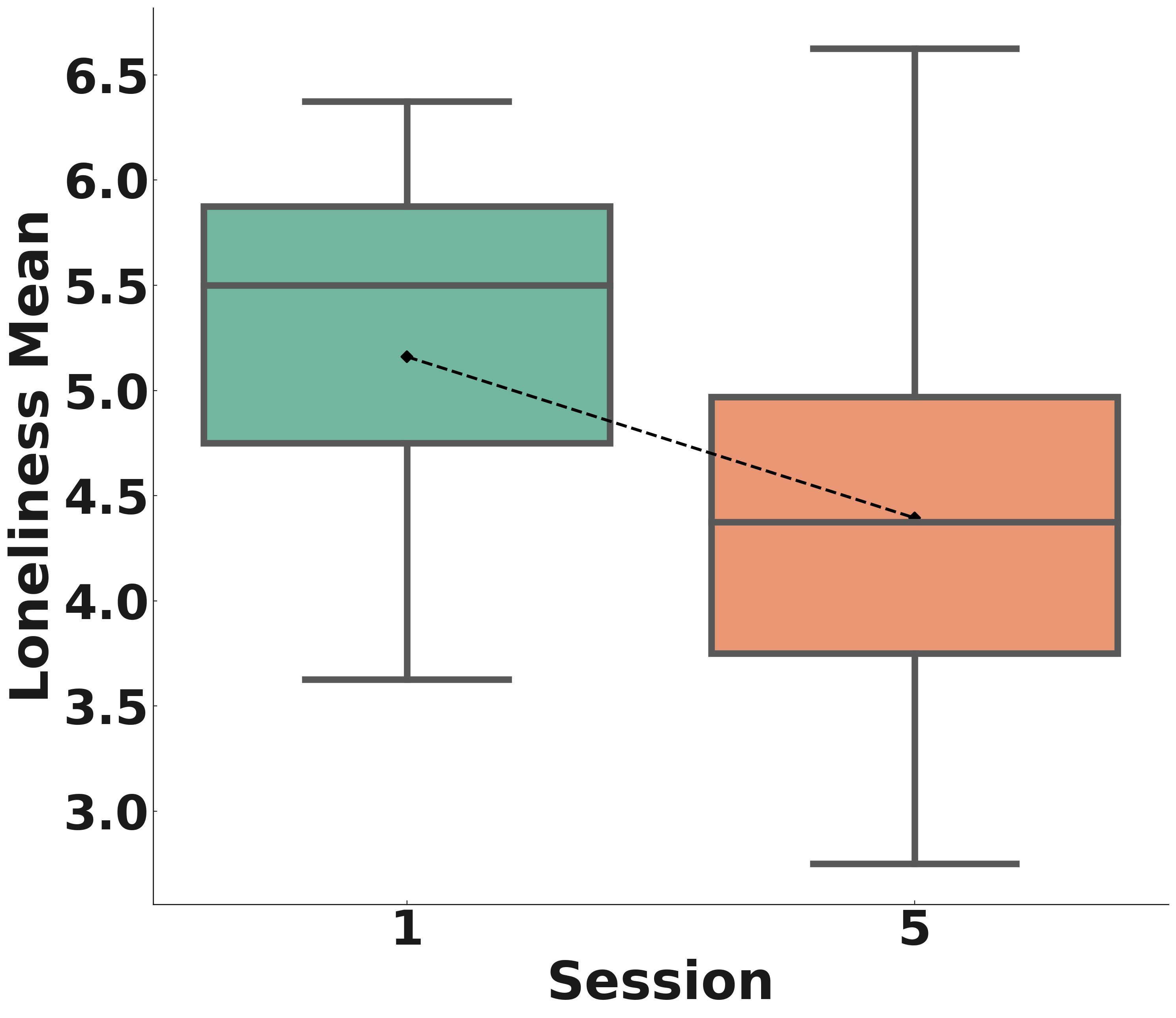}
    \includegraphics[width=0.49\linewidth]{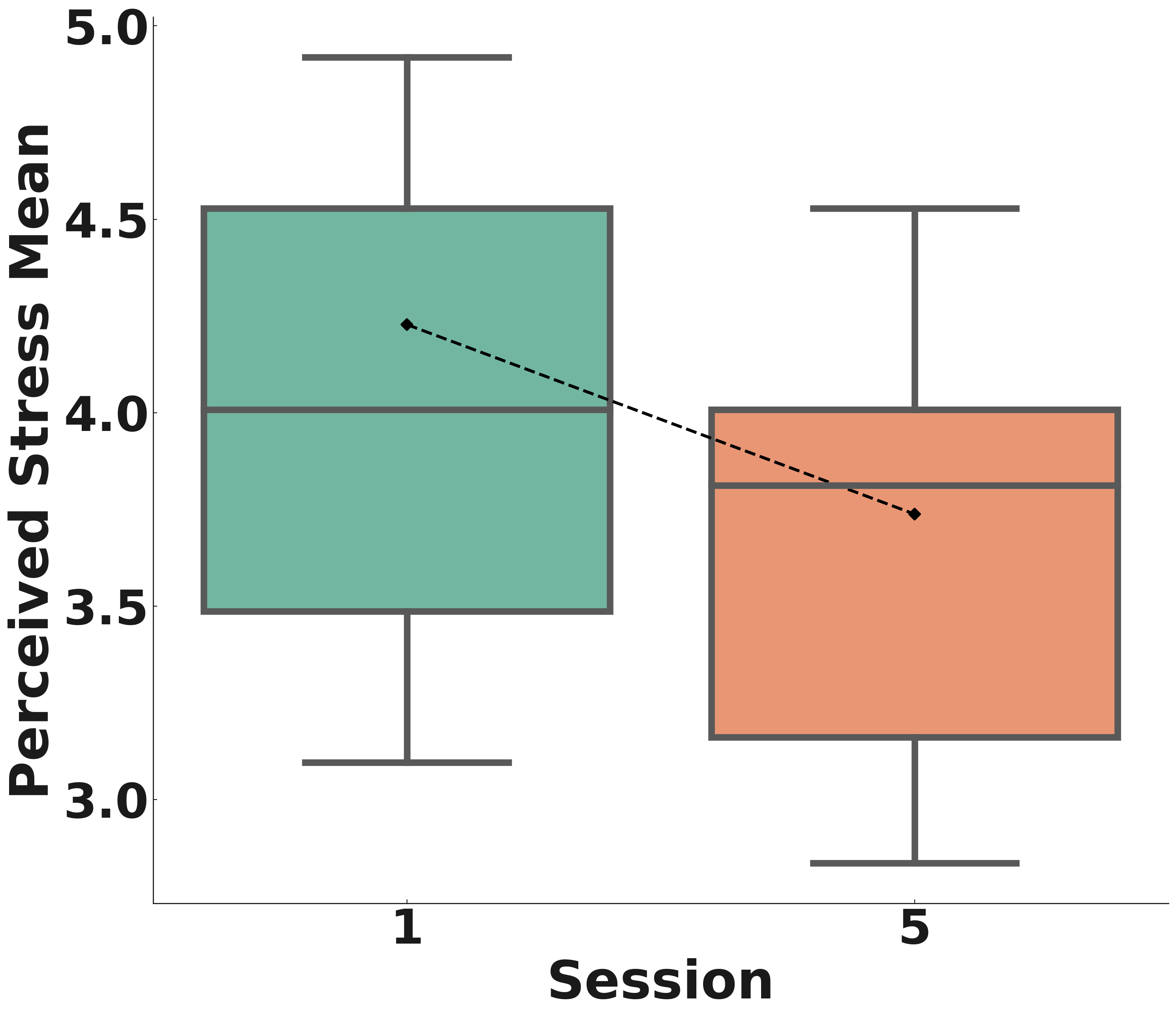}
    \caption{From left to right: (1) Mean loneliness score by session. (2) Mean perceived stress scores by session.}
    \label{fig:pss}
\end{figure}

Participants’ perceived stress significantly decreased across the five sessions. As shown in Figure~\ref{fig:pss}, participants reported less stress over time, $\beta$ = –0.06, $SE$ = 0.01, $p$ $<$ .001 (see Table~\ref{tab:lon_pss}). Despite substantial between-participant variability (±0.99), the results indicate a meaningful and consistent reduction in perceived stress associated with repeated interactions with QT.

\begin{table}[htbp!]
\caption{\small Mixed Effects Results of Loneliness and Perceived Stress Scores}\label{tab:lon_pss}
\centering\resizebox{\columnwidth}{!}{%
\begin{tabular}{lcccc}
\toprule
 & \multicolumn{2}{c}{Loneliness} & \multicolumn{2}{c}{Perceived Stress} \\
\cmidrule(lr){2-3} \cmidrule(lr){4-5}
\textbf{Predictors} & \textbf{Estimates} & \textbf{95\% CI} & \textbf{Estimates} & \textbf{95\% CI} \\
\midrule
Intercept & 5.18*** & 4.74–5.63 & 4.24*** & 3.80–4.69 \\
Session   & –0.15*** & –0.17––0.13 & –0.06*** & –0.07––0.05 \\
\midrule
\multicolumn{5}{l}{\textbf{Random Effects}} \\
$\sigma^2$  & \multicolumn{2}{c}{0.22} & \multicolumn{2}{c}{0.08} \\
$\tau_{00}$     & \multicolumn{2}{c}{0.96} & \multicolumn{2}{c}{0.98} \\
ICC                    & \multicolumn{2}{c}{0.81} & \multicolumn{2}{c}{0.92} \\
N (Participants)       & \multicolumn{2}{c}{21} & \multicolumn{2}{c}{21} \\
Observations           & \multicolumn{2}{c}{753} & \multicolumn{2}{c}{753} \\
Marginal $R^2$ / Conditional $R^2$ & \multicolumn{2}{c}{0.069 / 0.829} & \multicolumn{2}{c}{0.030 / 0.945} \\
\midrule
\multicolumn{5}{l}{\footnotesize{* $p$$<$0.05, ** $p$$<$0.01, *** $p$$<$0.001}} \\
\bottomrule
\end{tabular}}
\end{table}
\subsection{Identification of Disclosure Themes}

We identified six meaningful clusters in the dataset, encompassing a total of 560 disclosures, that yielded a total within-cluster sum of squares of 392.89. Cluster 0 comprised 113 responses (20.18\%), Cluster 1 included 101 responses (18.04\%), Cluster 2 included 78 responses (13.93\%), Cluster 3 included 95 responses (16.96\%), Cluster 4 included 96 responses (17.14\%), and Cluster 5 included 77 responses (13.75\%). Based on the LLM interpretation of central responses, these clusters were labelled as (0) Continuous Personal Development and Self-Reflection, (1) Building Connections and Memorable Experiences, (2) Academic Ambition and Future Aspirations, (3) Navigating Interpersonal Connections and Emotional Management, (4) Passion for Learning and Creativity, and (5) Friendships: Connection and Loneliness\footnote{See the clusters' descriptions here - \href{https://osf.io/crbpt}{https://osf.io/crbpt}}. The assigned description for each cluster consistently showed the highest similarity to its corresponding centroid, confirming that the LLM-generated labels captured the semantic core of each cluster (see  \ref{table:combined_validation}). 

\begin{table}[t]
\centering
\caption{Similarity Scores Between LLM Generated Descriptions \& Centroids}
\label{table:combined_validation}

\begin{tabular}{|c|c|c|c|c|c|c|}
\hline
\textbf{Cluster} & \textbf{Des. 0} & \textbf{Des. 1} & \textbf{Des. 2} & \textbf{Des. 3} & \textbf{Des. 4} & \textbf{Des. 5} \\
\hline
0 & \textbf{0.6088} & 0.4045 & 0.4971 & 0.4319 & 0.6126 & 0.2926 \\
1 & 0.2646 & \textbf{0.5031} & 0.2517 & 0.2243 & 0.3272 & 0.2890 \\
2 & 0.6067 & 0.3395 & \textbf{0.6262} & 0.3456 & 0.5552 & 0.2076 \\
3 & 0.4196 & 0.4181 & 0.2663 & \textbf{0.5816} & 0.4316 & 0.4812 \\
4 & 0.4431 & 0.2762 & 0.3834 & 0.3092 & \textbf{0.5639} & 0.1889 \\
5 & 0.3235 & 0.5918 & 0.2140 & 0.5374 & 0.3523 & \textbf{0.6319} \\
\hline
\end{tabular}
\end{table}

\subsection{Disclosure Theme and Loneliness}

We found a significant effect of interaction topic on loneliness, $\chi^2$(4) = 18.79, $p$ = .002. Post-hoc comparisons using the Mann-Whitney U test showed that participants in Building Connections and Memorable Experiences (Cluster 1; $M$ = 5.32, $SD$ = 0.99) reported significantly higher loneliness than those in Continuous Personal Development and Self-Reflection (Cluster 0; $M$ = 4.53, $SD$ = 0.85; $Z$ = 4.49, $p$ $<$ .001), Academic Ambition and Future Aspirations (Cluster 2; $M$ = 4.54, $SD$ = 1.34; $Z$ = 2.62, $p$ = .009), Navigating Interpersonal Connections and Emotional Management ($M$ = 4.60, $SD$ = 1.24; $Z$ = 4.09, $p$ = $<$.001), and Passion for Learning and Creativity (Cluster 4; $M$ = 4.76, $SD$ = 1.19; $Z$ = 2.06, $p$ = .039). Additionally, participants in Friendships: Connection and Loneliness (Cluster 5; $M$ = 5.03, $SD$ = 1.14) also reported significantly higher loneliness than those in Continuous Personal Development and Self-Reflection (Cluster 0; $M$ = 4.53, $SD$ = 0.85; $Z$ = 2.13, $p$ = .033), and those in Academic Ambition and Future Aspirations (Cluster 2; $M$ = 4.54, $SD$ = 1.34; -2.02, $p$ = .044). These findings suggest that conversations centred around social bonding and shared experiences are associated with higher perceived loneliness, while introspective or achievement-oriented topics correspond to lower loneliness. See the full pairwise comparison results in Table \ref{tab:pairwise_full}.

\subsection{Disclosure Theme and Perceived Stress}

We found a significant effect of disclosure topic on stress levels, $\chi^2$(4) = 17.03, $p$ = .004. Participants in \textit{Building Connections and Memorable Experiences} (Cluster 1; $M$ = 4.04, $SD$ = 0.45) reported significantly higher stress than those in \textit{Continuous Personal Development and Self-Reflection} (Cluster 0; $M$ = 3.61, $SD$ = 0.91; $Z$ = 2.53, $p$ = .011) and \textit{Academic Ambition and Future Aspirations} (Cluster 2; $M$ = 3.61, $SD$ = 0.91; $Z$ = 3.88, $p$ $<$ .001). Moreover, participants in \textit{Friendships: Connection and Loneliness} (Cluster 5; $M$ = 3.89, $SD$ = 0.69) reported significantly higher stress than those in \textit{Academic Ambition and Future Aspirations} (Cluster 2; $M$ = 3.61, $SD$ = 0.91; $Z$ = -2.24, $p$ = .025), and those in \textit{Passion for Learning and Creativity} (Cluster 4; $M$ = 3.85, $SD$ = 0.71) reported higher stress than participants in \textit{Academic Ambition and Future Aspirations} (Cluster 2; $M$ = 3.61, $SD$ = 0.91; $Z$ = 2.51, $p$ = .012). These findings suggest that conversations centred around social connection, friendships, and personal expression are associated with higher perceived stress, while discussions focused on academic goals and self-reflection correspond to lower stress levels. See the full pairwise comparison results in Table \ref{tab:pairwise_full}.


\begin{table*}[h!]
\centering
\caption{Pairwise Mann-Whitney U Comparisons across Disclosure Topics for Loneliness and Perceived Stress}
\resizebox{\textwidth}{!}{
\begin{tabular}{|p{2.5cm}|p{3.6cm}|p{3.6cm}|c c|c c|c|c|}
\hline
\textbf{Outcome} & \textbf{Cluster A} & \textbf{Cluster B} & \textbf{M\textsubscript{A}} & \textbf{SD\textsubscript{A}} & \textbf{M\textsubscript{B}} & \textbf{SD\textsubscript{B}} & \textbf{Z} & \textbf{$p$} \\
\hline
\multirow{15}{*}{Loneliness} 
& Personal Development (0) & Connections (1) & 4.53 & 0.85 & 5.32 & 0.99 & 4.49 & $<$.001 \\
& Personal Development (0) & Academic Ambition (2) & 4.53 & 0.85 & 4.54 & 1.34 & -0.15 & .880 \\
& Personal Development (0) & Emotional Management (3) & 4.53 & 0.85 & 4.60 & 1.24 & 0.25 & .804 \\
& Personal Development (0) & Learning (4) & 4.53 & 0.85 & 4.76 & 1.19 & 0.71 & .480 \\
& Personal Development (0) & Friendships (5) & 4.53 & 0.85 & 5.03 & 1.14 & 2.13 & .033 \\
& Connections (1) & Academic Ambition (2) & 5.32 & 0.99 & 4.54 & 1.34 & 2.62 & .009 \\
& Connections (1) & Emotional Management (3) & 5.32 & 0.99 & 4.60 & 1.24 & 4.09 & $<$.001 \\
& Connections (1) & Learning (4) & 5.32 & 0.99 & 4.76 & 1.19 & 2.06 & .039 \\
& Connections (1) & Friendships (5) & 5.32 & 0.99 & 5.03 & 1.14 & -1.49 & .137 \\
& Academic Ambition (2) & Emotional Management (3) & 4.54 & 1.34 & 4.60 & 1.24 & 0.58 & .562 \\
& Academic Ambition (2) & Learning (4) & 4.54 & 1.34 & 4.76 & 1.19 & 0.91 & .362 \\
& Academic Ambition (2) & Friendships (5) & 4.54 & 1.34 & 5.03 & 1.14 & 2.02 & .044 \\
& Emotional Management (3) & Learning (4) & 4.60 & 1.24 & 4.76 & 1.19 & 0.30 & .767 \\
& Emotional Management (3) & Friendships (5) & 4.60 & 1.24 & 5.03 & 1.14 & 1.10 & .272 \\
& Learning (4) & Friendships (5) & 4.76 & 1.19 & 5.03 & 1.14 & 1.49 & .139 \\
\hline
\multirow{15}{*}{Perceived Stress} 
& Personal Development (0) & Connections (1) & 3.61 & 0.91 & 4.04 & 0.45 & 2.53 & .011 \\
& Personal Development (0) & Academic Ambition (2) & 3.61 & 0.91 & 3.61 & 0.91 & -1.80 & .071 \\
& Personal Development (0) & Emotional Management (3) & 3.61 & 0.91 & 3.78 & 1.01 & 1.02 & .310 \\
& Personal Development (0) & Learning (4) & 3.61 & 0.91 & 3.85 & 0.71 & 0.38 & .705 \\
& Personal Development (0) & Friendships (5) & 3.61 & 0.91 & 3.89 & 0.69 & 0.74 & .461 \\
& Connections (1) & Academic Ambition (2) & 4.04 & 0.45 & 3.61 & 0.91 & 3.88 & $<$.001 \\
& Connections (1) & Emotional Management (3) & 4.04 & 0.45 & 3.78 & 1.01 & 1.34 & .180 \\
& Connections (1) & Learning (4) & 4.04 & 0.45 & 3.85 & 0.71 & 2.51 & .012 \\
& Connections (1) & Friendships (5) & 4.04 & 0.45 & 3.89 & 0.69 & -0.03 & .978 \\
& Academic Ambition (2) & Emotional Management (3) & 3.61 & 0.91 & 3.78 & 1.01 & -0.19 & .847 \\
& Academic Ambition (2) & Learning (4) & 3.61 & 0.91 & 3.85 & 0.71 & -0.91 & .365 \\
& Academic Ambition (2) & Friendships (5) & 3.61 & 0.91 & 3.89 & 0.69 & -2.24 & .025 \\
& Emotional Management (3) & Learning (4) & 3.78 & 1.01 & 3.85 & 0.71 & 1.10 & .272 \\
& Emotional Management (3) & Friendships (5) & 3.78 & 1.01 & 3.89 & 0.69 & -0.19 & .847 \\
& Learning (4) & Friendships (5) & 3.85 & 0.71 & 3.89 & 0.69 & 0.91 & .365 \\
\hline
\end{tabular}
}
\label{tab:pairwise_full}
\end{table*}

\section{Discussion}

Our findings provide empirical support for the emotional benefits of repeated interactions with a social robot powered by a LLM. Participants who engaged in five sessions with QTrobot reported significant reductions in both loneliness and perceived stress. These results align with prior work suggesting that social robots can promote well-being through 
self disclosure \cite{Laban2024BuildingTime,laban_ced_2023,Laban2024SharingFeel}, and extend previous findings by demonstrating that such improvements persist across repeated interactions when the robot is enhanced with more adaptive conversational capabilities (e.g., LLM).

Moreover, our analysis of disclosure themes offers novel insight into the relationship between emotional states and conversational content. We identified six distinct themes that emerged across participant disclosures, ranging from introspective topics such as personal development and academic ambition to more socially focused topics like friendships and shared experiences. Importantly, the themes associated with higher loneliness and stress scores were those emphasizing social connection—particularly Building Connections and Memorable Experiences and Friendships: Connection and Loneliness. In contrast, disclosures oriented toward Personal Development, Academic Ambition, and Passion for Learning were associated with lower emotional burden. Interestingly, this finding may suggest that conversations about relationships and connection serve not as indicators of fulfilled social bonds, but perhaps of a sense of lack or desire.

This aligns with prior findings and the theoretical perspective suggesting that lonely individuals often seek to discuss topics that reflect their unmet social needs or longings \cite{Hawkley2010LonelinessMechanisms,Horowitz1979InterpersonalLonely}. When participants chose to talk about friendship or social moments, it may have signified a preoccupation with disconnection or a call for closeness. These findings underscore the nuanced interplay between emotional states and the topics participants choose to discuss with a social robot. Participants experiencing heightened loneliness and stress appeared more inclined to reflect on the absence or longing for social bonds, while those with lower emotional distress were more likely to engage in self-affirming, goal-oriented topics. 
This outlook supports a more expressive view of disclosure towards robots: rather than reflecting one’s current state, what people choose to talk about with a robot may represent aspirations, deficits, or compensatory efforts. In that sense, the robot may function less like a passive recipient of affect and more like a projection space for expressing underlying emotional gaps.

These findings imply that conversational themes cannot be taken at face value as direct indicators of well-being. Instead, socially intelligent robots should be designed to interpret disclosures contextually. For instance, a user frequently speaking about friendship might not be celebrating connection but rather signalling loneliness—thus requiring responses that offer emotional affirmation, empathy, or opportunities to simulate social connection. Conversely, self-reflective or future-oriented disclosures might indicate a relatively resilient emotional state, and the robot could engage in more challenging, growth-oriented dialogues.

Nevertheless, several limitations should be acknowledged. First, our sample consisted of university students in a relatively controlled environment, which may limit generalizability to other populations or naturalistic settings. Second, while the use of an LLM allowed for adaptive conversation, participants may still have perceived the robot's responses as constrained, and future work should explore how perceptions of authenticity and agency affect emotional outcomes. In future studies, it would be valuable to investigate how real-time topic adaptation and emotional mirroring by the robot influence user engagement and well-being. 

\section{Conclusions}

This study provides evidence for the potential of social robots, powered by LLMs, to meaningfully reduce loneliness and perceived stress through repeated interactions. Beyond addressing these benefits, our findings highlight the nuanced interplay between feelings of loneliness and stress and conversational content, showing that participants experiencing greater emotional distress (i.e., loneliness and perceived stress) were more likely to disclose themes centred around social connection. These insights suggest that the topics people choose to share with a robot may reflect not only current states but also unmet socio-emotional needs or aspirations. These insights can help guide the design of socio-emotionally aware robots that respond more thoughtfully to users’ concerns and support them in meaningful ways.


\bibliographystyle{myIEEEtran2}
\balance{\bibliography{references-2,newbib}}

\begin{thebibliography}{10}
\providecommand{\url}[1]{#1}
\csname url@samestyle\endcsname
\providecommand{\newblock}{\relax}
\providecommand{\bibinfo}[2]{#2}
\providecommand{\BIBentrySTDinterwordspacing}{\spaceskip=0pt\relax}
\providecommand{\BIBentryALTinterwordstretchfactor}{4}
\providecommand{\BIBentryALTinterwordspacing}{\spaceskip=\fontdimen2\font plus
\BIBentryALTinterwordstretchfactor\fontdimen3\font minus \fontdimen4\font\relax}
\providecommand{\BIBforeignlanguage}[2]{{%
\expandafter\ifx\csname l@#1\endcsname\relax
\typeout{** WARNING: IEEEtran.bst: No hyphenation pattern has been}%
\typeout{** loaded for the language `#1'. Using the pattern for}%
\typeout{** the default language instead.}%
\else
\language=\csname l@#1\endcsname
\fi
#2}}
\providecommand{\BIBdecl}{\relax}
\BIBdecl

\bibitem{Arnett2014TheHealth}
J.~J. Arnett, R.~{\v{Z}}ukauskiene, and K.~Sugimura, ``{The new life stage of emerging adulthood at ages 18-29 years: Implications for mental health},'' \emph{The Lancet Psychiatry}, vol.~1, no.~7, pp. 569--576, 12 2014.

\bibitem{Shah2023UnderstandingInterventions.}
H.~A. Shah and M.~Househ, ``{Understanding Loneliness in Younger People: Review of the Opportunities and Challenges for Loneliness Interventions.}'' \emph{Interactive journal of medical research}, vol.~12, no.~1, p. e45197, 11 2023.

\bibitem{Magid2023TheStudy}
K.~Magid, S.~J. Sagui-Henson, C.~C. Sweet, B.~J. Smith, C.~E. Chamberlain, and S.~M. Levens, ``{The Impact of Digital Mental Health Services on Loneliness and Mental Health: Results from a Prospective, Observational Study},'' \emph{International Journal of Behavioral Medicine}, vol.~31, no.~3, p. 468, 6 2023.

\bibitem{Ellard2023Review:Review}
O.~B. Ellard, C.~Dennison, and H.~Tuomainen, ``{Review: Interventions addressing loneliness amongst university students: a systematic review},'' \emph{Child and Adolescent Mental Health}, vol.~28, no.~4, pp. 512--523, 11 2023.

\bibitem{Masi2010ALoneliness}
C.~M. Masi, H.-Y. Chen, L.~C. Hawkley, and J.~T. Cacioppo, ``{A Meta-Analysis of Interventions to Reduce Loneliness},'' \emph{Personality and Social Psychology Review}, vol.~15, no.~3, pp. 219--266, 8 2010.

\bibitem{breazeal_designing_2004}
C.~L. Breazeal, \emph{{Designing sociable robots}}.\hskip 1em plus 0.5em minus 0.4em\relax MIT press, 2004.

\bibitem{Laban2024SocialWell-Being}
G.~Laban, V.~Morrison, and E.~Cross, ``{Social Robots for Health Psychology: A New Frontier for Improving Human Health and Well-Being},'' \emph{European Health Psychologist}, vol.~23, no.~1, pp. 1095--1102, 2 2024.

\bibitem{Henschel2021}
A.~Henschel, G.~Laban, and E.~S. Cross, ``{What Makes a Robot Social? A Review of Social Robots from Science Fiction to a Home or Hospital Near You},'' \emph{Current Robotics Reports}, no.~2, pp. 9--19, 2021.

\bibitem{RefWorks:404}
N.~L. Robinson, T.~V. Cottier, and D.~J. Kavanagh, ``{Psychosocial Health Interventions by Social Robots: Systematic Review of Randomized Controlled Trials},'' \emph{J Med Internet Res}, vol.~21, no.~5, pp. 1--20, 2019.

\bibitem{Laban2024SharingFeel}
G.~Laban and E.~S. Cross, ``{Sharing our Emotions with Robots: Why do we do it and how does it make us feel?}'' \emph{IEEE Transactions on Affective Computing}, pp. 1--18, 2024.

\bibitem{Laban2021TellSpeech}
G.~Laban, J.-N. George, V.~Morrison, and E.~S. Cross, ``Tell me more! assessing interactions with social robots from speech,'' \emph{Paladyn, Journal of Behavioral Robotics}, vol.~12, no.~1, pp. 136--159, 2021.

\bibitem{Langer2019}
A.~Langer, R.~Feingold-Polak, O.~Mueller, P.~Kellmeyer, and S.~Levy-Tzedek, ``{Trust in socially assistive robots: Considerations for use in rehabilitation},'' pp. 231--239, 9 2019.

\bibitem{Stower}
R.~Stower, N.~Calvo-Barajas, G.~Castellano, and A.~Kappas, ``A meta-analysis on children’s trust in social robots,'' \emph{International Journal of Social Robotics}, vol.~13, pp. 1979--2001, 12 2021.

\bibitem{Laban2024BuildingTime}
G.~Laban, A.~Kappas, V.~Morrison, and E.~S. Cross, ``{Building Long-Term Human–Robot Relationships: Examining Disclosure, Perception and Well-Being Across Time},'' \emph{International Journal of Social Robotics}, vol.~16, no.~5, pp. 1--27, 2024.

\bibitem{laban_ced_2023}
G.~Laban, V.~Morrison, A.~Kappas, and E.~S. Cross, ``{Coping with Emotional Distress via Self-Disclosure to Robots: An Intervention with Caregivers},'' \emph{International Journal of Social Robotics}, 2025.

\bibitem{SpitaleMicol2025VITA:Coaching}
M.~Spitale, M.~Axelsson, and H.~Gunes, ``{VITA: A Multi-Modal LLM-Based System for Longitudinal, Autonomous and Adaptive Robotic Mental Well-Being Coaching},'' \emph{ACM Transactions on Human-Robot Interaction}, vol.~14, no.~2, pp. 1--28, 3 2025.

\bibitem{Laban2025AReappraisal}
G.~Laban, J.~Wang, and H.~Gunes, ``{A Robot-Led Intervention for Emotion Regulation: From Expression to Reappraisal},'' 3 2025.

\bibitem{Cacioppo2006LonelinessPerspective}
J.~T. Cacioppo, L.~C. Hawkley, J.~M. Ernst, M.~Burleson, G.~G. Berntson, B.~Nouriani, and D.~Spiegel, ``{Loneliness within a nomological net: An evolutionary perspective},'' \emph{Journal of Research in Personality}, vol.~40, no.~6, pp. 1054--1085, 12 2006.

\bibitem{Hawkley2010LonelinessMechanisms}
L.~C. Hawkley and J.~T. Cacioppo, ``{Loneliness Matters: A Theoretical and Empirical Review of Consequences and Mechanisms},'' \emph{Annals of behavioral medicine : a publication of the Society of Behavioral Medicine}, vol.~40, no.~2, pp. 218--227, 10 2010.

\bibitem{lazarus1984stress}
R.~S. Lazarus and S.~Folkman, \emph{{Stress, appraisal, and coping}}.\hskip 1em plus 0.5em minus 0.4em\relax Springer publishing company, 1984.

\bibitem{Delgado2023CharacterizingConnection}
M.~R. Delgado, D.~S. Fareri, and L.~J. Chang, ``{Characterizing the mechanisms of social connection},'' \emph{Neuron}, vol. 111, no.~24, pp. 3911--3925, 12 2023.

\bibitem{Horowitz1979InterpersonalLonely}
L.~M. Horowitz and R.~de~Sales~French, ``{Interpersonal problems of people who describe themselves as lonely},'' \emph{Journal of Consulting and Clinical Psychology}, vol.~47, no.~4, pp. 762--764, 8 1979.

\bibitem{Pennebaker1993PuttingImplications}
J.~W. Pennebaker, ``{Putting stress into words: Health, linguistic, and therapeutic implications},'' \emph{Behaviour Research and Therapy}, vol.~31, no.~6, pp. 539--548, 7 1993.

\bibitem{Cacioppo2014SocialIsolation}
J.~T. Cacioppo and S.~Cacioppo, ``{Social Relationships and Health: The Toxic Effects of Perceived Social Isolation},'' \emph{Social and Personality Psychology Compass}, vol.~8, no.~2, pp. 58--72, 2 2014.

\bibitem{Cacioppo2011SocialIsolation}
J.~T. Cacioppo, L.~C. Hawkley, G.~J. Norman, and G.~G. Berntson, ``{Social isolation},'' \emph{Annals of the New York Academy of Sciences}, vol. 1231, no.~1, pp. 17--22, 8 2011.

\bibitem{Segrin2007PositiveWell-being}
C.~Segrin and M.~Taylor, ``{Positive interpersonal relationships mediate the association between social skills and psychological well-being},'' \emph{Personality and Individual Differences}, vol.~43, no.~4, pp. 637--646, 9 2007.

\bibitem{Segrin2010FunctionsHealth}
C.~Segrin and S.~A. Passalacqua, ``{Functions of Loneliness, Social Support, Health Behaviors, and Stress in Association With Poor Health},'' \emph{Health Communication}, vol.~25, no.~4, pp. 312--322, 6 2010.

\bibitem{Zaki2013InterpersonalRegulation}
J.~Zaki and C.~W. Williams, ``{Interpersonal emotion regulation},'' \emph{Emotion}, vol.~13, no.~5, pp. 803--810, 10 2013.

\bibitem{Zaki2020IntegratingRegulation}
J.~Zaki, ``{Integrating Empathy and Interpersonal Emotion Regulation},'' \emph{Annual Review of Psychology}, vol.~71, pp. 517--540, 1 2020.

\bibitem{Gasteiger2021FriendsPeople}
N.~Gasteiger, K.~Loveys, M.~Law, and E.~Broadbent, ``{Friends from the future: A scoping review of research into robots and computer agents to combat loneliness in older people},'' \emph{Clinical Interventions in Aging}, vol.~16, pp. 941--971, 2021.

\bibitem{Rasouli2022PotentialAnxiety}
S.~Rasouli, G.~Gupta, E.~Nilsen, and K.~Dautenhahn, ``{Potential Applications of Social Robots in Robot-Assisted Interventions for Social Anxiety},'' \emph{International Journal of Social Robotics 2022 14:5}, vol.~14, no.~5, pp. 1--32, 1 2022.

\bibitem{Laban2022SocialTreatmentd}
G.~Laban, Z.~Ben-Zion, and E.~S. Cross, ``{Social Robots for Supporting Post-traumatic Stress Disorder Diagnosis and Treatment},'' \emph{Frontiers in psychiatry}, vol.~12, 2022.

\bibitem{2023OpeningBehavior}
G.~Laban, A.~Kappas, V.~Morrison, and E.~S. Cross, ``{Opening Up to Social Robots: How Emotions Drive Self-Disclosure Behavior},'' in \emph{2023 32nd IEEE International Conference on Robot and Human Interactive Communication (RO-MAN)}, 8 2023, pp. 1697--1704.

\bibitem{penner2022}
A.~Penner and F.~Eyssel, ``{Germ-Free Robotic Friends: Loneliness during the COVID-19 Pandemic Enhanced the Willingness to Self-Disclose towards Robots},'' \emph{Robotics}, vol.~11, no.~6, 2022.

\bibitem{RefWorks:300}
N.~Epley, A.~Waytz, S.~Akalis, and J.~T. Cacioppo, ``{When We Need A Human: Motivational Determinants of Anthropomorphism},'' \emph{Social Cognition}, vol.~26, no.~2, pp. 143--155, 2008.

\bibitem{Epley2008CreatingArticle}
N.~Epley, S.~Akalis, A.~Waytz, and J.~T. Cacioppo, ``{Creating social connection through inferential reproduction: Loneliness and perceived agency in gadgets, gods, and hreyhounds: Research article},'' \emph{Psychological Science}, vol.~19, no.~2, pp. 114--120, 2 2008.

\bibitem{Yamaguchi2023YoungStress}
M.~Yamaguchi, M.~Okanda, Y.~Moriguchi, and S.~Itakura, ``{Young adults with imaginary companions: The role of anthropomorphism, loneliness, and perceived stress},'' \emph{Personality and Individual Differences}, vol. 207, p. 112159, 6 2023.

\bibitem{Leichtmann2025LonelyRobots}
B.~Leichtmann, E.~Gollob, M.~May, A.~Paschmanns, and M.~Mara, ``{Lonely Minds and Robotic Bonds: Effects of Human Loneliness on the Anthropomorphization of Robots},'' \emph{International Journal of Social Robotics 2025}, pp. 1--27, 3 2025.

\bibitem{10715872}
M.-Y. Lin and et~al., ``Embodied ai with large language models: A survey and new hri framework,'' in \emph{2024 International Conference on Advanced Robotics and Mechatronics (ICARM)}, 2024, pp. 978--983.

\bibitem{Seligman2018PERMAWell-being}
M.~E.~P. Seligman, ``{PERMA and the building blocks of well-being},'' \emph{The Journal of Positive Psychology}, vol.~13, no.~4, pp. 333--335, 7 2018.

\bibitem{Hays1987}
R.~D. Hays and M.~R. DiMatteo, ``{A Short-Form Measure of Loneliness},'' \emph{Journal of Personality Assessment}, vol.~51, no.~1, pp. 69--81, 3 1987.

\bibitem{RefWorks:475}
S.~Cohen, T.~Kamarck, and R.~Mermelstein, ``{A global measure of perceived stress},'' \emph{Journal of health and social behavior}, vol.~24, no.~4, pp. 385--396, 1983.

\bibitem{Bates2015FittingLme4}
D.~Bates, M.~M{\"{a}}chler, B.~M. Bolker, and S.~C. Walker, ``{Fitting Linear Mixed-Effects Models Using lme4},'' \emph{Journal of Statistical Software}, vol.~67, no.~1, pp. 1--48, 10 2015.

\bibitem{Bell2019FixedChoice}
A.~Bell, M.~Fairbrother, and K.~Jones, ``{Fixed and random effects models: making an informed choice},'' \emph{Quality and Quantity}, vol.~53, no.~2, pp. 1051--1074, 3 2019.

\bibitem{Hesser2015ModelingInterventions}
H.~Hesser, ``{Modeling individual differences in randomized experiments using growth models: Recommendations for design, statistical analysis and reporting of results of internet interventions},'' \emph{Internet Interventions}, vol.~2, no.~2, pp. 110--120, 5 2015.

\bibitem{Kuznetsova2017LmerTestModels}
A.~Kuznetsova, P.~B. Brockhoff, and R.~H. Christensen, ``{lmerTest Package: Tests in Linear Mixed Effects Models},'' \emph{Journal of Statistical Software}, vol.~82, no.~13, pp. 1--26, 12 2017.

\bibitem{Satterthwaite1946AnComponents}
F.~E. Satterthwaite, ``{An Approximate Distribution of Estimates of Variance Components},'' \emph{Biometrics Bulletin}, vol.~2, no.~6, p. 110, 12 1946.

\bibitem{10.5555/3495724.3496209}
W.~Wang, F.~Wei, L.~Dong, H.~Bao, N.~Yang, and M.~Zhou, ``Minilm: deep self-attention distillation for task-agnostic compression of pre-trained transformers,'' in \emph{Proceedings of the 34th Int. Conference on Neural Information Processing Systems}, 2020.

\bibitem{Reimers2019Sentence-BERT:BERT-Networks}
N.~Reimers and I.~Gurevych, ``{Sentence-BERT: Sentence Embeddings using Siamese BERT-Networks},'' \emph{Proceedings of EMNLP-IJCNLP 2019}, pp. 3982--3992, 2019.

\bibitem{likas2003global}
A.~Likas, N.~Vlassis, and J.~J. Verbeek, ``The global k-means clustering algorithm,'' \emph{Pattern recognition}, vol.~36, no.~2, pp. 451--461, 2003.

\bibitem{liu2020determine}
F.~Liu and Y.~Deng, ``Determine the number of unknown targets in open world based on elbow method,'' \emph{IEEE Transactions on Fuzzy Systems}, vol.~29, no.~5, pp. 986--995, 2020.

\end{thebibliography}


\end{document}